\pdfoutput=1
\documentclass[prl,a4paper,amsfonts,amssymb,amsmath,reprint,
showkeys,twoside,superscriptaddress,longbibliography]{revtex4-2}
\usepackage[english]{babel}
\usepackage[utf8]{inputenc}
\usepackage[colorinlistoftodos, color=green!40, prependcaption]{todonotes}
\usepackage{amsthm}
\usepackage{mathtools}
\usepackage{physics}
\usepackage{xcolor}
\usepackage{graphicx}
\usepackage[normalem]{ulem}
\usepackage[left=23mm,right=13mm,top=35mm,columnsep=15pt]{geometry} 
\usepackage{adjustbox}
\usepackage{placeins}
\usepackage[T1]{fontenc}
\usepackage{lipsum}
\usepackage{csquotes}

\usepackage[pdftex, pdftitle={Article}, pdfauthor={Author}]{hyperref} 
\newcommand{\sub}[2]{{#1}_{\mbox{\!\! \scriptsize #2}}}

\newcommand{\bv}[1]{\mathbf{ #1 }}

\def\beq{\begin{equation}}
\def\eeq{\end{equation}}

\def\CR{\nonumber\\[0.15cm]}

\newcommand{\fref}[1]{Fig.~\ref{#1}}
\newcommand{\frefp}[2]{Fig.~\ref{#1}~(#2)}

\newcommand{\eref}[1]{Eq.~(\ref{#1})}

\newcommand{\cref}[1]{chapter~\ref{#1}}

\newcommand{\Cref}[1]{Chapter~\ref{#1}}

\begin{document}
\title{Low-Energy Purification of Crystal Defects by Rydberg Excitons}

\author{Shiva Kant Tiwari}
\affiliation{Department of Chemistry, Purdue University, West Lafayette, Indiana 47907, USA}
\author{Tijs Karman}
\affiliation{Institute for Molecules and Materials, Radboud University, 6525 AJ Nijmegen, Netherlands}
\author{Valentin Walther}
\email{vwalther@purdue.edu}
\affiliation{Department of Physics and Astronomy, Purdue University, West Lafayette, Indiana 47907, USA}
\affiliation{Purdue Quantum Science and Engineering Institute, Purdue University, West Lafayette, Indiana 47907, USA}
\affiliation{Department of Chemistry, Purdue University, West Lafayette, Indiana 47907, USA}

\date{\today}

\begin{abstract}
Recent experiments show that optically generated Rydberg excitons in cuprous oxide (Cu$_2$O) can neutralize charged impurities, strongly reducing stray electric fields and effectively “purifying” the crystal. Here, we develop a multichannel theory of Rydberg exciton–impurity scattering that resolves the competing roles of capture, elastic scattering, and inelastic transitions between excitonic states.  
We find that at high collision energies, as effective under conventional single-photon excitation, purification is reduced relative to Langevin capture. These collisions are accompanied by inelastic redistribution and dominant elastic scattering, including pronounced glory scattering, which suppress purification efficiency. 
We identify a quantum regime at ultralow collision energies favorable for purification, where only the $s$-wave contributes: capture is enhanced while elastic and inelastic channels are strongly suppressed. This regime can be accessed via degenerate two-photon excitation of even-parity Rydberg excitons with tunable recoil, additionally enabling the systematic exploration of exciton–impurity scattering over a wide range of collision energies beyond what is readily achievable in atomic counterparts in atomic gas experiments.
\end{abstract}

	\maketitle
%

%
\begin{figure}[h]
    \centering
    \includegraphics[width=1.0\linewidth]{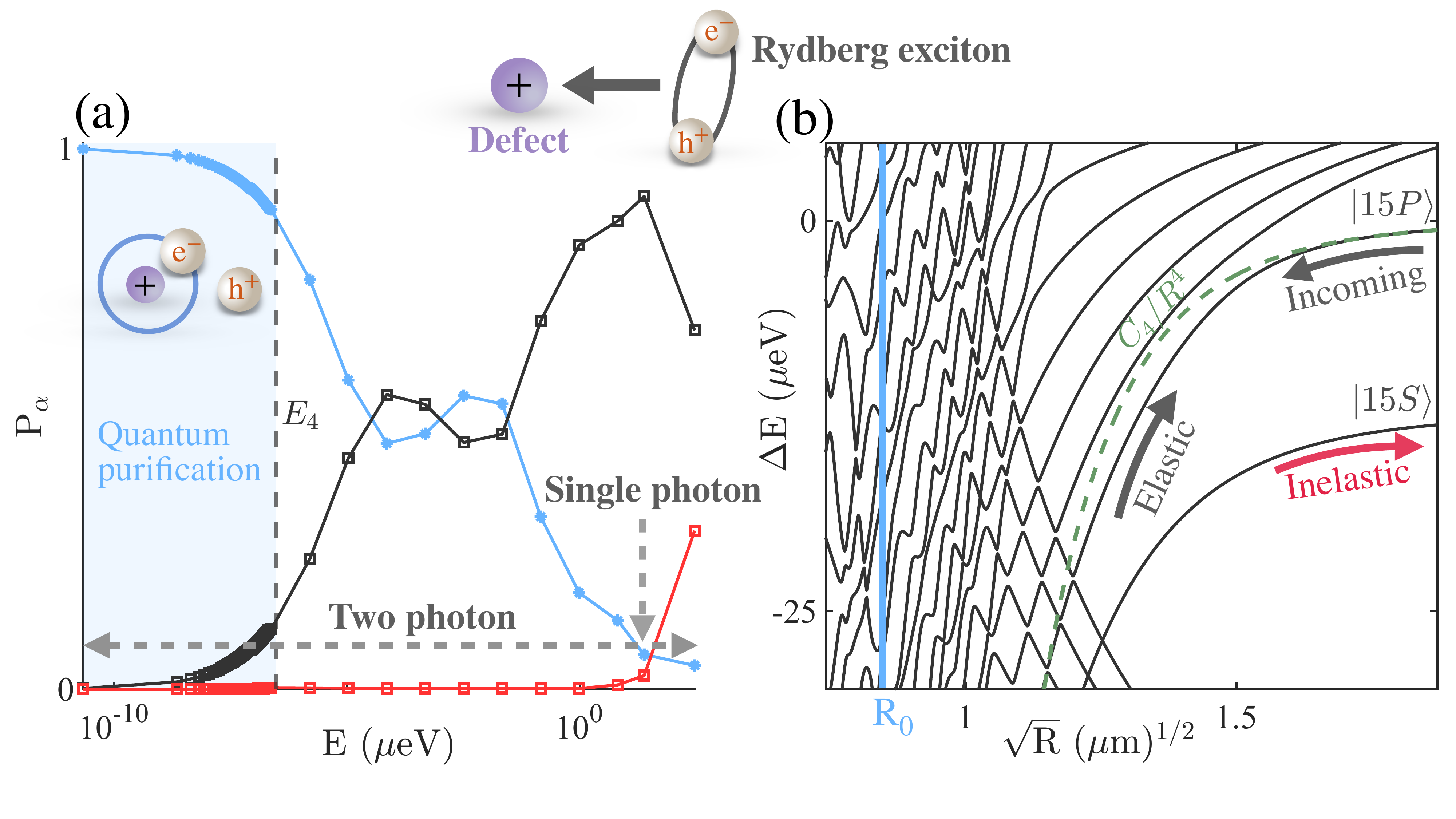}
    \caption{\textbf{Energy-dependent scattering channels and adiabatic potentials:} (a) Schematic illustration of a Rydberg exciton (gray ellipse) approaching a charged defect (blue), with the arrow indicating the direction of motion. Over the recoil-energy range accessible by the two-photon excitation scheme, the collision can lead to three competing outcomes: elastic scattering, inelastic scattering (with a change of internal state), and reactive capture (schematically shown as capture of the electron by the defect, leaving the hole unbound). The energy-dependent branching fractions are defined as $P_{\alpha}(E)=\sigma_{\alpha}(E)/\sigma_{\mathrm{tot}}(E)$, with $\alpha\in\{\mathrm{reactive},\mathrm{elastic},\mathrm{inelastic}\}$. Reactive capture dominates at low energies (blue line), whereas elastic scattering (black line) becomes increasingly important at higher energies, accompanied by a finite inelastic contribution (red line). (b) Adiabatic potential-energy surfaces (solid black lines) of the Rydberg exciton--defect interaction, referenced to the asymptotic energy of the incoming Rydberg state $\ket{15P}$. The polarization potential (dashed red line) is obtained by fitting the long-range behavior of the $\ket{15P}$ surface, yielding $C_4 = 8.8\times10^{-5}\,\mathrm{eV\,\mu m^4}$ and the associated natural scales $R_4 \approx 42~\mu\mathrm{m}$ and $E_4 \approx 0.3~\mathrm{peV}$. For the even-parity $\ket{15S}$ Rydberg state, a slightly different $C_4$ gives $R_4 \approx 47~\mu\mathrm{m}$ and $E_4 \approx 0.2~\mathrm{peV}$. The reactive boundary condition is imposed at $R_0 = 0.73~\mu\mathrm{m}$ (vertical blue line).
} 
    \label{fig:Sketch}
\end{figure}

The quality of semiconductors is often restricted by the presence of defects that limit the coherence of transport and lifetimes of electronic dynamics. One very important type of defects are charged vacancy centers that emerge in crystal growth as well as in natural crystals~\cite{basset2014evaluating,nguyen2011impurity,kestner2013noise,barnes2011screening}. 
Their charges generate spatially fluctuating electric fields that induce random Stark shifts and spectral broadening, thereby limiting coherence, reproducibility, and control of electron motion and spin dynamics in semiconductors \cite{kuhlmann2013charge,hu2006charge,culcer2009dephasing}. Strategies to eliminate the effects of charged impurities include the introduction of free charge carriers, either by gating or by illuminating the sample above the band-gap, that neutralize the defects \cite{houel2012probing,hauck2014locating}, evening out the stray disordered electric fields. However, the presence of free charge carriers can lead to other, often undesired, effects, such as trion~\cite{PhysRevB.92.205418,PhysRevB.96.035131,PhysRevB.96.085302,PhysRevB.101.195417,PhysRevB.105.L041302,PhysRevB.109.085406,PhysRevLett.132.036903} or polaron formation~\cite{sidler2017fermi,PhysRevB.95.035417}. It would therefore be highly desirable to find a minimally invasive approach to disable the effect of impurities. 

The impact of charge defect is especially pronounced for Rydberg-state excitons, whose large polarizability renders them extremely sensitive to local electric fields \cite{kruger2020interaction}. At the same time, the strong Rydberg exciton–defect interaction provides a natural mechanism to neutralize charged defects, as was recently demonstrated in bulk cuprous oxide (Cu$_2$O) \cite{bergen2023large}. These internally excited states of an electron-hole pair are charge neutral, can be selectively excited to principal quantum numbers of $n\sim 30$ \cite{kazimierczuk2014giant,versteegh2021giant} and exhibit exaggerated properties known from Rydberg atoms, including enormous spatial extents of up to a micron and long lifetimes \cite{chakrabarti2025direct}, as well as giant optical response \cite{zielinska2026propagation, brewin2024microwave, PhysRevLett.131.033607, PhysRevB.105.075307}. These excitons can collide with the defects, such that the electron (hole) is captured by a positively (negatively) charged impurity, while the hole (electron) is freed up and can be expelled to the crystal surface. The neutralization of localized charges is known as "purification". The precise capture rates can be chosen by optical selection of the Rydberg state's internal quantum state, with a demonstrated $\sim n^{3.5}$ scaling of the capture cross section \cite{bergen2023large}.

The purification process has previously been described within a simplified classical picture in \cite{bergen2023large}. The incoming excitons are understood to be shot into motion by the photon recoil of the excitation laser, after which they collide classically with a simple point charge. This description effectively captures the observed scaling of the capture cross section with the principal quantum number. However, it overlooks central features of Rydberg excitons, namely the internal complexity of excitonic states, which may open up inelastic scattering channels where the excitons, instead of neutralizing the impurity, exit in a different internal quantum state. Furthermore, it remains to be understood how excitons scatter when created under optical excitation conditions that impart much less recoil momentum, where the classical description breaks down.

In this Letter, we develop a comprehensive theory of scattering between Rydberg excitons and point defects. We first show that the underlying quantum theory predicts reduced capture relative to the classical capture picture and reveals competing inelastic and elastic scattering channels.
We then identify an experimentally accessible low-energy regime in which scattering becomes fully quantum ($s$-wave dominated) and show that the purification is strongly enhanced, whereas competing channels are suppressed.

The scattering of Rydberg excitons with defects can be captured by the following model: Excitonic Rydberg states have positronium-like wave functions of an effective conduction band electron of mass $m_e$ and a valence band hole of mass $m_h$, bound together by a screened Coulomb interaction with relative permittivity $\epsilon_r$. The defect states are modeled as fixed points of charge $Q$ that exert a potential on the excitons. Through this potential, the excitonic states are polarized, and experience a polarization force at large distances, whereas higher-order interactions and state mixing become relevant at short interactions.

While many of the conclusions are general to Rydberg excitons across various semiconductor platforms, we focus here on bulk Cu$_2$O, whose excitons have been observed at record principal quantum numbers, and where purification has already been demonstrated experimentally
\cite{bergen2023large}. In Cu$_2$O, defect-related photoluminescence studies identified singly and doubly ionized oxygen vacancies, $V_{\mathrm{O}}^{+}$ and $V_{\mathrm{O}}^{2+}$, as well as copper vacancies, with oxygen-vacancy-related emission being stronger in natural crystals~\cite{lynch2021rydberg}. However, a recent first-principles study suggests interstitials over vacancies~\cite{brewin2026cu2o}. At distances of more than a few lattice constants ($\approx 0.427$ nm \cite{lynch2021rydberg}), much shorter than the exciton size, the potential of these defects is well described by an isotropic Coulomb potential with relative permittivity $\epsilon_r = 7.5$. The excitonic wave functions and energies are computed under the inclusion of non-parabolic dispersions of the band structure, that yield a Rydberg series of internal states with Rydberg constant of $R_y = 86\, \mathrm{meV}$ and effective quantum defects \cite{kruger2020interaction}, as well as free center of mass dynamics with a total mass $M = 1.75\, \mathrm{m_e}$ \cite{hodby1976cyclotron,naka2012time}. Compared to Rydberg atoms, Rydberg excitons in Cu$_2$O are thus much lighter and much less bound, making interaction effects more pronounced overall.

To achieve a systematic and physically controlled separation between the universal long-range interaction and short-range excitonic structure, we perform a multipole expansion of the exciton–ion interaction (Sec.~S1 \cite{SM}).
The resulting potential energy surfaces (PESs), shown in Fig.~\ref{fig:Sketch}(b), encode the short-range structure generated by the multipole couplings and asymptotically approach the universal long-range polarization potential. At large separations, the interaction is dominated by the charge–induced dipole potential, $V(R) = -C_4/R^4$, where the coefficients $C_4$ are set by the exciton polarizability $\alpha$ and $R$ is the exciton separation from the impurity. For Rydberg excitons, the polarizability scales as $\alpha \propto \alpha_0 n^7$~\cite{heckotter2017scaling,hahn2000polarizability,saffman2010quantum}, with $\alpha_0$ the ground-state polarizability, leading to strongly enhanced long-range interactions at high principal quantum numbers. Similar shapes of the PESs in the Rydberg atom–ion interaction lead to the formation of long-range Rydberg–ion molecules~\cite{PhysRevResearch.3.023114}, which are absent here due to the much lighter exciton mass.

A first rough understanding of the general scattering dynamics can be obtained from the simplified picture of a single potential energy curve of the asymptotic $-C_4/R^4$ polarization potential shape. 
This potential defines polarization length and energy scales, $R_{4}=\sqrt{M C_4}/\hbar$ and $E_4=\hbar^4/(M^2C_4)$. Physically, $R_4$ is the distance beyond which the exciton behaves approximately as a free particle, while $E_4$ marks the crossover, for an incoming exciton of energy $E$, from quantum ($E \lesssim E_4$: $s$-wave) to semiclassical ($E \gtrsim E_4$) dynamics. 
Compared to atom--ion systems \cite{kamenski2014formal}, the polarization coefficient is dramatically larger for excitons because of their enhanced polarizability. However, as a consequence of the smaller masses, the associated exciton--impurity polarization length scale is about an order of magnitude smaller, while the corresponding energy scale is roughly $10^{6}$ times larger, making the quantum regime far more accessible.

Despite this, the first purification experiments in Cu$_2$O operated at collision energies in the $\mu\mathrm{eV}$ range, which far exceed $E_4$ and made the collisions essentially semiclassical. At these energies, many de Broglie wavelengths fit within $R_4$ ($kR_4\approx 10^{3}$, where $k=\sqrt{2\,M \,E}/\hbar$ is the wave number), decohering quantum interference effects.  
Such collisional energies are reached from the recoil energy in single-laser excitation of the Rydberg state. 
In this limit, the polarization potential significantly affects the motion only at micron distances, $R\le R_c^*\approx 1.4~\mu\mathrm{m}$, estimated from $V(R_c^*)\sim E$. Thus, under the conditions of previous experiments, the dynamics is dominated by short-range physics. 
In this semiclassical limit, the short-range dynamics can be interpreted as classical capture~\cite{bergen2023large}, as formulated in the Langevin model~\cite{cote2000ultracold}: trajectories with impact parameter below a critical value surmount the centrifugal barrier, reach short range, and are irreversibly lost, yielding a loss cross section $\sigma_{L}\propto E^{-1/2}$, see Sec.~S5 \cite{SM} for a detailed discussion. 
To test this prediction, we solve the dynamics on a single adiabatic potential energy surface obtained by fitting the computed microscopic potentials (\frefp{fig:Sketch}{b}), with absorbing boundary conditions at short range. Indeed, we find recover the Langevin prediction for the energy scaling and the effective partial-wave cutoff (Fig.~1 Sec.~S3 \cite{SM}).

However, this simplified single-potential model neglects the complex structure of the PESs (Fig.~\ref{fig:Sketch}{b}) that could redistribute incoming flux over neighboring manifolds. To address this question, a full multichannel scattering treatment is required, opening pathways to inelastic processes beyond pure capture and state-preserving scattering.

We formulate the dynamics on the adiabatic PESs shown in Fig.~(\ref{fig:Sketch}{b}). Under a local gauge transformation (Sec.~S2 \cite{SM}) that transfers the nonadiabatic couplings to the diabatic potential matrix ($V_{ab}(R)$), one obtains the coupled radial Schr\"odinger equation governing the relative motion. The motion on adiabatic potential energy surface $a$ for partial wave $l$ is then governed by
\begin{align}
\label{eq:radial}
    &\left[-\frac{\hbar^2}{2M}\frac{d^2}{dR^2} 
    + \frac{\hbar^2 l(l+1)}{2M R^2} \right] u^l_{a}(R)
    + \sum_b V_{ab}(R)\,u^l_{b}(R)\CR
    &= E\,u^l_{a}(R).
\end{align}
 Equation~\eqref{eq:radial} forms the basis of our scattering calculations. We solve the coupled radial equations for every partial-wave order $l$ numerically using the renormalized Numerov method~\cite{karman2014renormalized}. 

The channel-resolved cross sections for different scattering processes are directly obtained from the $S$~matrix \cite{thompson2009nuclear} (see Sec.~S3 \cite{SM}).
We distinguish between the reactive ($\sigma_{\mathrm{re}}^{(a)}$), inelastic ($\sigma_{\mathrm{in}}^{(a)}$), and elastic ($\sigma_{\mathrm{el}}^{(a)}$) cross sections from the incoming channel $a$, respectively.
Together, the three state-resolved cross sections give a complete physical picture of the Rydberg exciton scattering with the impurity. 
The scattering processes are computed using an absorbing boundary condition at short range. The results are insensitive to the placement of the absorbing boundary once the cutoff $R_0$ is chosen within the converged regime (see, Sec.~S3 \cite{SM}). As shown in Fig.~\ref{fig:Sketch}(a), the relative weights of reactive, elastic, and inelastic scattering vary strongly with $E$, reflecting the changing accessibility of short range and multichannel flux redistribution; a detailed discussion is given below.

\begin{figure}[h!]
    \centering
    \includegraphics[width=1.0\columnwidth]{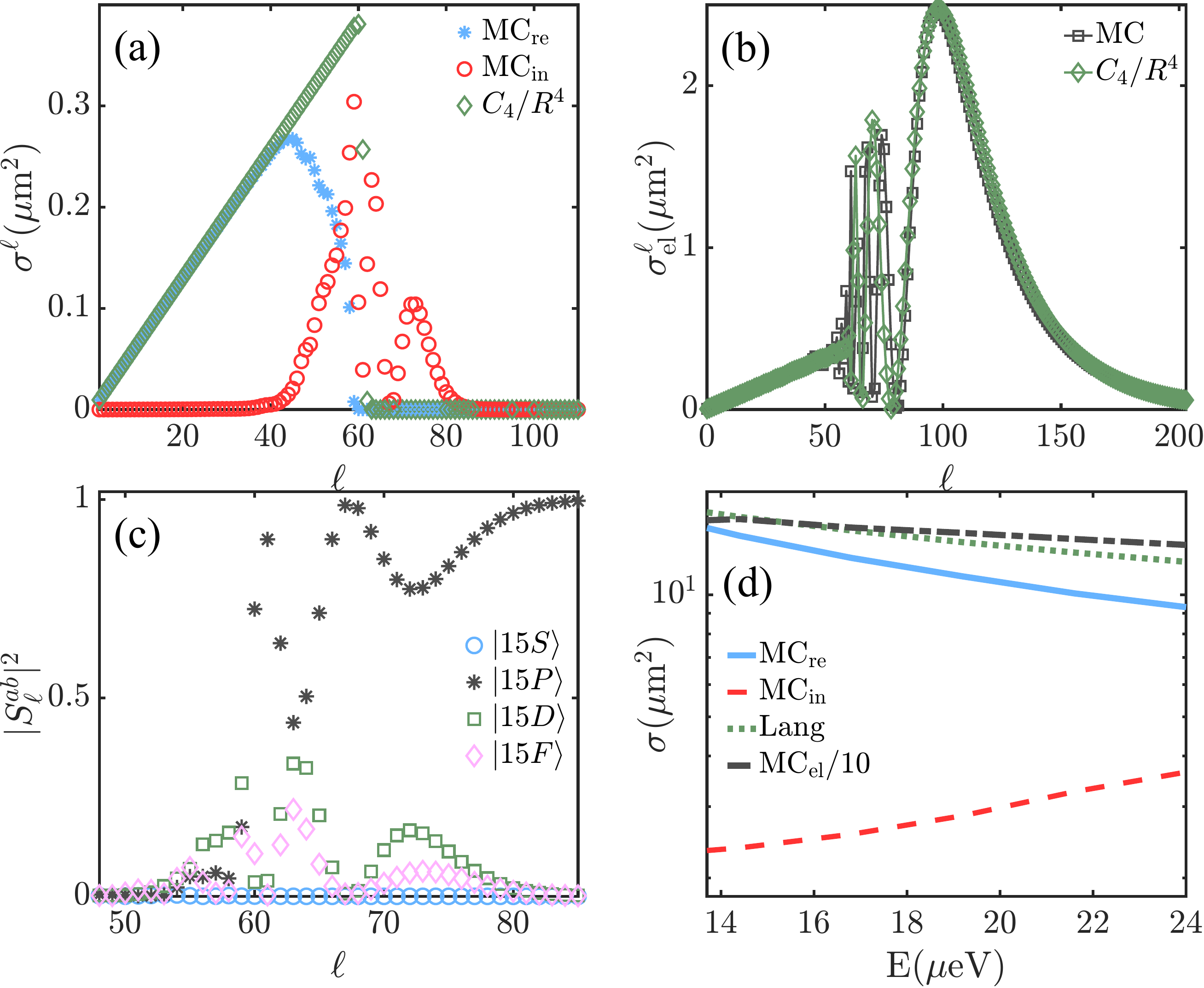}
    \caption{\textbf{Semiclassical partial-wave dynamics:} (a) The partial-wave reactive cross-section for the multichannel case (blue $\star$) decreases smoothly due to the onset of inelastic scattering (red $\circ$), whereas for the single-channel polarization potential (green $\diamond$) it drops sharply beyond $l \approx 60$ . (b) The elastic cross-section for the polarization potential (black $\square$, solid line) shows deviations from the multichannel result (green $\diamond$, solid line) in the intermediate-$l$ regime. (c) Population initially in the $\ket{15P}$ state (blue $\star$) is predominantly transferred to the $\ket{15D}$ (green $\square$) and $\ket{15F}$ (magenta $\diamond$) manifolds, while transfer to the $\ket{15S}$ state (blue $\circ$) is negligible. (a)–(c) are shown for $E = 24~\mu\mathrm{eV}$. (d) The total reactive cross-section (solid blue line) is reduced relative to the Langevin prediction (dotted green line) due to inelastic scattering (dashed red line), while the total elastic cross-section (dash-dotted black line) exceeds the Langevin prediction by approximately an order of magnitude. The elastic cross-section is scaled by $10^{-1}$ for visibility.}
    \label{fig:PartialReactive}
\end{figure}

We begin our analysis in the high-energy limit of Fig.~\ref{fig:Sketch}(a). In this regime, the scattering is controlled by the competition between the collision energy and the centrifugal barrier. For low $l$, the collision energy exceeds the centrifugal barrier, so the exciton penetrates to short range and is fully captured ($|S^{aa}_l|^2 \approx 0$), see Sec.~S4 \cite{SM}. As a result, exciton--impurity scattering exhibits a large reactive cross section that grows linearly with $l$, similar to the polarization-potential case, see \frefp{fig:PartialReactive}{a}. Similar behavior is observed for the elastic cross section, as shown in Fig.~\ref{fig:PartialReactive}(b). Despite a large oscillating phase shift, the elastic cross section also increases linearly with $l$ due to complete capture of Rydberg exciton. 
As $l$ increases, the centrifugal barrier progressively becomes comparable to the collision energy, reducing short-range access. Capture becomes less efficient, and the missing flux is redistributed among nearby excitonic states, as exemplified in Fig.~\ref{fig:PartialReactive}(c). In this region, the imparted phase shift oscillates rapidly and is effectively random modulo $\pi$, satisfies the random-phase approximation ($\langle \sin^2 \delta_l \rangle \approx 1/2$)~\cite{levine1987molecular} and produces oscillations in the multichannel elastic cross section, distinct from the polarization-potential result. For sufficiently large $l$, the centrifugal barrier suppresses short-range interactions altogether, and scattering is governed solely by the long-range polarization potential. The reactive and inelastic contributions vanish, and the elastic cross section smoothly decreases because the phase shift decays as $\delta_l \propto E/l^3$, consistent with the eikonal approximation~\cite{landau422lifshitz}. Because inelastic redistribution occurs over a finite range of partial waves, the total reactive cross section is reduced relative to the Langevin prediction, as shown in Fig.~\ref{fig:PartialReactive}(d). Meanwhile, the total elastic cross section exceeds the total loss by nearly an order of magnitude and follows the semiclassical scaling $\sigma_{\mathrm{el}}\propto E^{-1/3}$~\cite{rios2020introduction}, i.e., exhibits a weaker decay with growing energy than Langevin loss. 

\begin{figure}[h!]
    \centering
    \includegraphics[width=1.0\columnwidth]{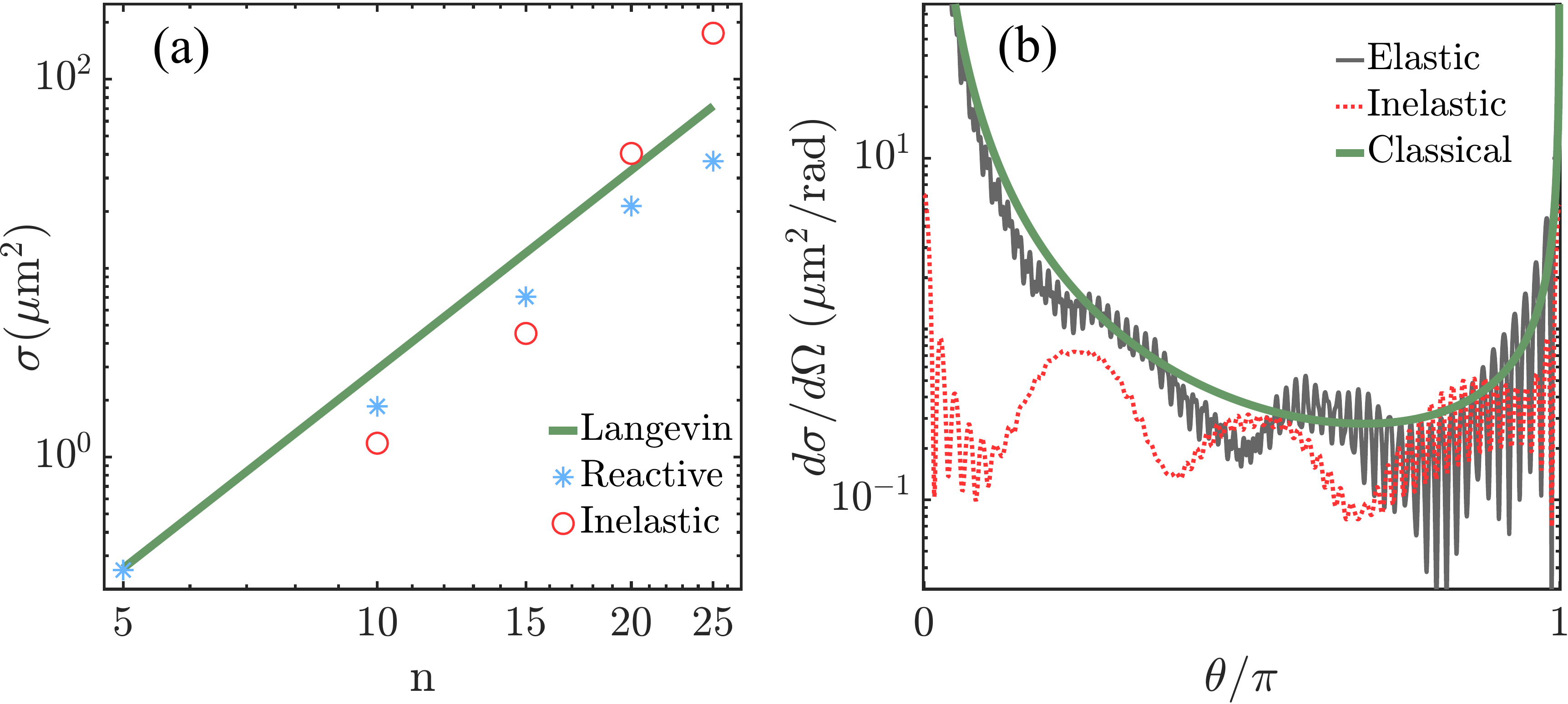}
    \caption{\textbf{Cross-section scaling and angular distributions:} (a) The reactive contribution (blue $\star$) exhibits a reduced scaling, $\sigma_{\mathrm{re}}\propto n^{3.1}$, relative to the Langevin prediction (solid green line), owing to inelastic scattering (blue $\circ$). The $n=5$ inelastic point lies below the plot scale, with $\sigma_{\mathrm{in}} = 7.2\times10^{-5}~\mu\mathrm{m}^2$. (b) The elastic differential cross section (solid black) exhibits strong forward and backward peaks with prominent interference fringes. Its envelope is well described by the classical result (solid green) and is clearly distinct from the inelastic contribution (dotted red).
 }
    \label{fig:DiffCross}
\end{figure}

%
In the high-energy regime, scattering includes both capture and state-changing pathways. Since the interaction strength grows with principal quantum number as $C_4\propto n^{7}$, the Langevin model predicts a total loss scaling $\sigma_{L}\propto n^{3.5}$.
Our multichannel calculations, shown in Fig.~\ref{fig:DiffCross}(a), demonstrate that the reactive component alone exhibits a slightly weaker scaling with $n$. However, we find that when total loss is defined to include both reactive and inelastic contributions from the partial waves within the polarization-induced capture region, the Langevin scaling is recovered. This suggests that, in a multichannel setting, the classical Langevin prediction effectively accounts for all flux entering the long-range loss region, irrespective of whether it leads to capture or inelastic redistribution. By contrast, additional nonadiabatic coupling outside this loss region (Sec.~S4 \cite{SM}) leads to deviations from the Langevin scaling, particularly at larger principal quantum numbers. While both reactive and inelastic processes contribute to the total loss in theory, only the reactive channel corresponds to true removal of exciton from the initial state and is therefore directly measurable as purification. In contrast, inelastic processes populate nearby Rydberg states and require additional state resolution to be observed. Consequently, tuning $n$ provides an experimental handle to disentangle purification from inelastic redistribution.

Angle-resolved differential cross sections of the outgoing excitonic channels in Fig.~\ref{fig:DiffCross}(b) provide an alternative route to disentangle collision pathways. The elastic signal exhibits interference fringes with strong forward and backward enhancements, arising from coherent superposition of many partial waves. Its envelope follows the classical polarization-potential result, with a forward divergence $d\sigma_{\mathrm{el}}/d\Omega \propto \theta^{-5/2}$ set by large impact parameters (see Sec.~S5~\cite{SM}), while the backward peak and oscillations reflect glory scattering \cite{adam2002mathematical,berry1972semiclassical} (see Sec.~S6~\cite{SM}). In contrast, the inelastic contribution exhibits a distinct interference pattern reflecting coherence among state-changing scattering pathways. Here, the reactive contribution can be inferred from the missing flux after integrating the elastic and inelastic channels.

%


An important open question is how the collision energy affects the efficiency of the purification process. 
At sufficiently low collision energies (``quantum purification'' in \frefp{fig:Sketch}{a}), exciton--impurity scattering enters a distinct regime in which only a few partial waves contribute, thereby enabling controlled quantum dynamics. To access it, we propose a two-photon excitation scheme that resonantly excites even-parity Rydberg excitons \cite{heckotter2025energy, PhysRevB.105.115206}. The idea is that two counter-propagating beams yield a recoil momentum $\mathbf{K}=\mathbf{k}_1+\mathbf{k}_2$, and resulting recoil energy $E = \hbar^2 |\mathbf{K}|^2/(2M)$ which can be controlled both through the laser frequencies and the relative angle between the excitation beams. If slightly detuned lasers are used, this excitation scheme yields essentially background-free excitation of Rydberg states at ultralow energies.

\begin{figure}[h!]
    \centering
    \includegraphics[width=1.0\columnwidth]{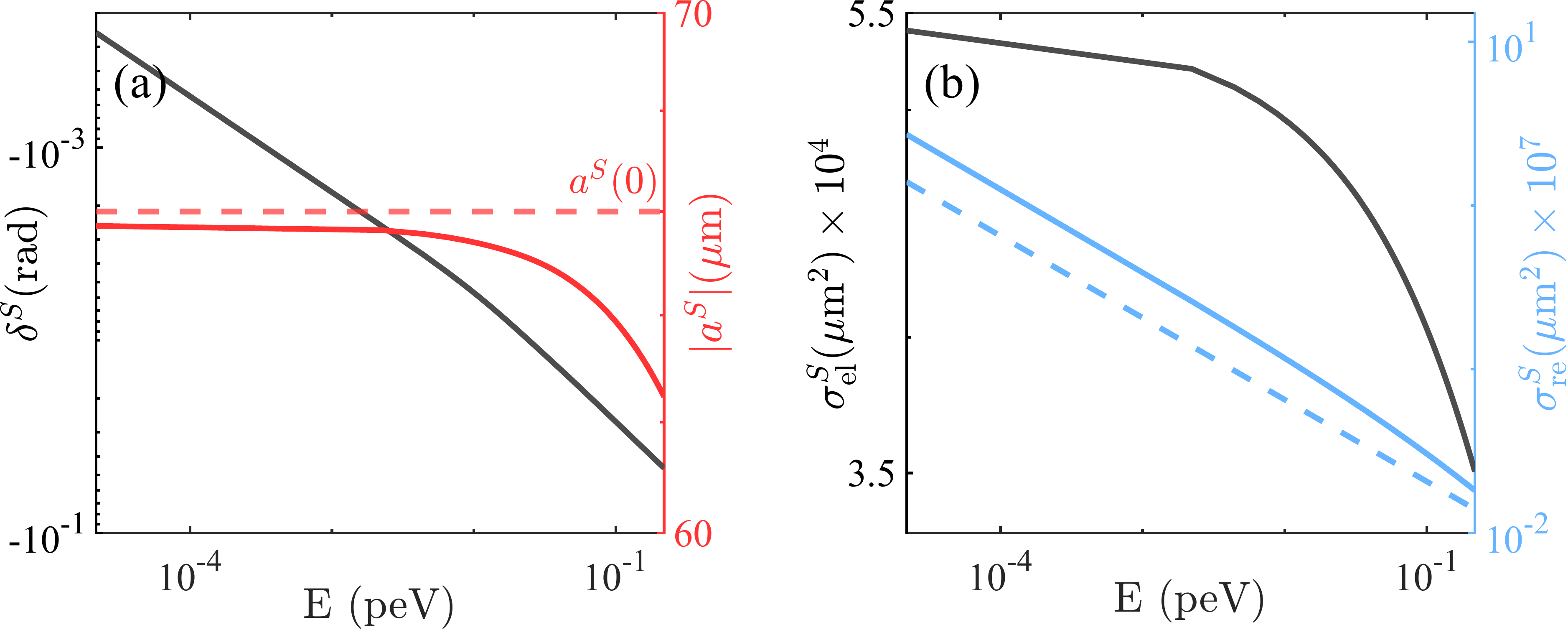}
    \caption{\textbf{Quantum-enhanced purification:} (a) Phase shifts (solid black line) obtained from the multichannel calculation decrease as the collision energy is reduced. The scattering length (solid red line) converges to the analytical zero-energy value ($\approx 66 \mathrm{\mu m}$) for the polarization potential (dashed red line). (b) The elastic cross section (solid black line) saturates to a constant value as $E \to 0$. The reactive cross section (solid blue line) diverges at low energy, following an $\propto E^{-1/2}$ scaling and is twice the Langevin prediction (dashed blue line).}
    \label{fig:LowPhase}
\end{figure}

In the ultra-low-energy regime, 
the dimensionless wavenumber is $kR_4 \ll 1$, placing the collision deep in the $s$-wave limit. The dynamics is then controlled by large separations, with a characteristic interaction radius $R_c^* \gtrsim 10^{2}~\mu\mathrm{m}$, so nonadiabatic multichannel effects are strongly suppressed and scattering is dominated by the polarization tail. In the $s$-wave regime, the short-range radial wavefunction for the polarization potential has the form $\exp\!\bigl(-i\sqrt{2}\,R_4/R\bigr)$, and the low-energy phase shift $\delta^{S}(k)$ vanishes linearly with the collision momentum $k$~\cite{idziaszek2011multichannel}. The corresponding zero-energy scattering length is $a^{S}(0) = -i\sqrt{2}\,R_4$\cite{PhysRevA.91.030701,PhysRevA.48.546,PhysRevLett.110.213202}. Consistently, our multichannel calculations show $\delta^{S}\to 0$ linearly as $E\to 0$ (Fig.~\ref{fig:LowPhase}(a)), and the extracted scattering length approaches that of the polarization-potential.

In this regime, where the scattering is fully characterized by the $s$-wave scattering length, the behavior of the elastic and reactive cross-sections follows directly from the zero-energy solution. In particular, the elastic cross-section saturates to a constant value $\sigma^{S}_{\mathrm{el}} = 8\pi\, (R_4)^2$, as shown in Fig.~\ref{fig:LowPhase}(b). This saturation arises because both the incoming rate, $(\hbar\, k\,)/M$ and scattered rates, $(8\pi\, \hbar\, k\, (R_4)^2)/M$ vanish linearly with energy, resulting in a finite ratio and hence a constant elastic cross-section. In contrast, the total loss rate, $(4 \sqrt{2}\,\pi\,\hbar\, R_4)/M$ remains finite as $E \to 0$, leading to a diverging reactive cross-section $\sigma^{S}_{\mathrm{re}} = 4\sqrt{2}\,\pi\,R_4/k$.  
The reactive cross-section in this low energy regime ($E_{\mathrm{low}}$) is approximately $2\sqrt{E_{\rm high}/E_{\rm low}}$ times larger than that at collision energies ($E_{\mathrm{high}}$) under single-photon excitation, indicating a dramatic enhancement of excitonic purification efficiency. At the same, the suppression of inelastic processes in this limit provides a favorable regime for achieving highly efficient exciton purification. Finally, it is interesting to note that the reactive cross-sections approaches a value that is twice that of the Langevin prediction~\cite{vogt1954scattering,fabrikant2001low} (Figure~\ref{fig:LowPhase}(b)). This quantum enhancement is a reflection of the wave nature of the exciton, whose wavefunction is larger than the defect \cite{landau2013quantum}.

In conclusion, we establish that the crossover from semiclassical to quantum scattering is controlled by the collision energy, which can be tuned experimentally through the exciton recoil momentum. At high collision energies, we find reduced purification relative to the Langevin model used to interpret earlier experiments, accompanied by inelastic and elastic scattering as well as the emergence of backward glory scattering. In this regime, short-range physics is essential and can be incorporated systematically. Here we model capture via an absorbing boundary condition, which can be refined to include microscopic short-range dynamics, including three-body processes \cite{braaten2006universality} and exciton breakup during purification.  In contrast, at low collision energies only the $s$-wave contributes; elastic and inelastic channels are strongly suppressed while reactive capture is enhanced, identifying a quantum regime for highly efficient exciton-driven purification of cuprous oxide.
The predicted crossover from semiclassical Langevin behavior to the quantum threshold regime in Rydberg exciton–defect collisions should also exist for Rydberg excitons in two-dimensional materials \cite{2DMaterials}, as well as in atom–ion collisions.  
Unlike atomic systems, where collision energies are limited by trapping temperatures \cite{Tilman;PRL}, excitonic scattering energies can be tuned over orders of magnitude through optical excitation and recoil control. This tunability, together with the multichannel structure of Rydberg excitons, establishes them as as a promising solid-state platform for exploring quantum scattering in long-range interacting systems, with  Cu$_2$O quantum wells offering an appealing direction for future studies in confined geometries \cite{1yqx-syzy}. 

\begin{acknowledgments}
\textit{Acknowledgments.---}
This work has primarily been supported by the National Science Foundation under Award $\#$PHYS-2409630. VW also acknowledges support from the U.S. Department of Energy (DOE), Office of Science, Basic Energy Sciences (BES), under Award $\#$DE-SC0026318.
\end{acknowledgments}

%

\clearpage

\onecolumngrid

\appendix

\section*{Supplemental Material}

\setcounter{section}{0}
\renewcommand{\thesection}{S\arabic{section}}

\setcounter{equation}{0}
\renewcommand{\theequation}{S\arabic{equation}}

\setcounter{figure}{0}
\renewcommand{\thefigure}{S\arabic{figure}}

\setcounter{table}{0}
\renewcommand{\thetable}{S\arabic{table}}

\makeatletter
\@addtoreset{equation}{section}
\makeatother
	\subsection{S1.~ Rydberg Exciton--Ion Interaction Potential} \label{sec;RydPot}

%
We begin by constructing the interaction between a Rydberg exciton and a defect in cuprous oxide (${\rm Cu}_2{\rm O}$) through a multipole expansion of the defect potential. This yields a controlled long-range multichannel interaction while retaining the essential internal quantum structure of the exciton. In cuprous oxide, Coulomb interactions are screened by the background dielectric constant, $\epsilon_r\simeq 7.5$ for ${\rm Cu}_2{\rm O}$~\cite{hodby1976cyclotron}.
In the presence of an charged impurity, the exciton--impurity interaction potential is
\begin{align}
\sub{V}{I}(\bv{R},\bv{r})
= -\frac{e^2}{4 \pi \epsilon_0\epsilon_r} \bigg(-\frac{1}{|\alpha \bv{r}+\bv{R}|} + \frac{1}{|\beta \bv{r}-\bv{R}|} \bigg),
\end{align}
where $\alpha = m_h/M$ and $\beta = m_e/M$ denote the mass ratios of the hole and electron, respectively, with respect to the total exciton mass $M = m_e + m_h \approx 1.57\,m_e$. The vector $\bv{R} = \sub{\bv{R}}{cm} - \sub{\bv{R}}{ion}$ denotes the displacement between the exciton center of mass and the defect, and $\bv{r} = \bv{r}_e - \bv{r}_h$ is the electron--hole relative coordinate. To extract the leading behavior at large separations, we perform a multipole expansion in the regime $r \ll R$, yielding
\begin{align} \label{Multi}
\sub{V}{I}(\bv{R},\bv{r})
&= -\frac{e^2}{4 \pi \epsilon_0\epsilon_r} \sum_{\lambda=1}^{\infty} \sqrt{\frac{4\pi}{2\lambda +1}} \frac{r^\lambda}{R^{\lambda+1}}
\,[\beta^{\lambda} -(-\alpha)^{\lambda}]\, Y_{\lambda 0}(\hat{\bv{r}}),
\end{align}
where $Y_{\lambda 0}$ are spherical harmonics and the polar axis is taken along $\bv{R}$, so that the multipole expansion is cylindrically symmetric about the exciton–impurity axis and only the $m_{\lambda}=0$ components appear. Here $\lambda$ is the multipole order, corresponding to charge-dipole ($\lambda = 1$), charge-quadrupole ($\lambda=2$), and higher moments.

In contrast to the multipole expansion for Rydberg atom--ion interactions, where the atomic core is typically treated as infinitely massive, the exciton--impurity potential retains an explicit mass dependence. In the limit $m_h \to \infty$, the expansion reduces to the standard Rydberg atom--ion form~\cite{Tilman;PRL}. Throughout, the impurity is treated as a fixed external center, and impurity recoil is neglected.  

The leading long-range contribution of \eref{Multi} is the charge--induced-dipole interaction. The impurity field induces a dipole moment at first order by mixing opposite-parity exciton states, while the associated energy shift arises at second order as the familiar Stark shift
\begin{align}\label{StarkS}
    \Delta E(R) = -\frac{C_4}{R^4}.
\end{align}
Here, $C_4$ is determined by the polarizability (see main text), and we adopt the convention $V(R)=-C_4/R^4$.
%
\section{S2.~ Coupled-Channel Schr\"odinger Equation for Exciton--Ion Scattering}\label{sec;RadEqu}
%
With the multipole expansion of the Rydberg-exciton--impurity interaction, Eq.~\eqref{Multi}, in center-of-mass and relative coordinates, the total Hamiltonian reads
\begin{align}
   H = -\frac{\hbar^2}{2 M}\bm{\nabla}^2_{\bv{R}} + H_{X}(\bv{r}) + \sub{V}{I}(\bv{R},\bv{r}),
\end{align}
Here, the operator $H_X(\bv{r})$ is the bare Hamiltonian of the Rydberg exciton including Coulomb interaction and the quantum defects~\cite{heckotter2017scaling}. 
The time-independent Schr\"odinger equation for the total wavefunction $\psi$ is
\begin{align}
    H\,\psi(\bv{R},\bv{r}) = E\,\psi(\bv{R},\bv{r}),
\end{align}
where $E$ is the collision energy set by the recoil energy of the optically generated Rydberg exciton. For each fixed $R=|\bv{R}|$, the adiabatic excitonic states and potential energy surfaces are obtained from
\begin{align}
    \bigl[H_{X}(\bv{r}) + \sub{V}{I}(\bv{R},\bv{r})\bigr]\chi_{a}(R;\bv{r})
    = \epsilon_{a}(R)\,\chi_{a}(R;\bv{r}).
\end{align}
Here $\chi_a$ denotes the adiabatic excitonic states and $\epsilon_a(R)$ their corresponding potential energy surfaces (PES), shown as solid black lines in Fig.~1(b) of the main article. These adiabatic potentials are obtained by progressively including additional excitonic states and higher-order multipole terms in the interaction until convergence of the relevant observables is achieved. For all results presented in the main text, convergence is reached when three excitonic manifolds (principal quantum numbers) above and below the target state are included and the multipole expansion is retained up to seventh order. No restriction is made on the remaining quantum numbers. As seen in Fig.~1(b) of the main article, the adiabatic surfaces asymptotically approach polarization potentials in the long-range limit. Expanding the total wavefunction as
\begin{align}
    \psi(\bv{R},\bv{r}) = \sum_{a}\phi_{a}(\bv{R})\,\chi_{a}(R;\bv{r}),
\end{align}
one obtains the coupled Schr\"odinger equation in the adiabatic representation~\cite{domcke2004conical},
\begin{align}\label{ExcitonEqua1}
     -\frac{\hbar^2}{2\mu}\bigl(\boldsymbol{\nabla}_{\bv{R}}+\mathbf{F}(\bv{R})\bigr)^2 \boldsymbol{\phi}(\bv{R})
     + \boldsymbol{\epsilon}(R)\,\boldsymbol{\phi}(\bv{R})
     = E\,\boldsymbol{\phi}(\bv{R}),
\end{align}
where $\boldsymbol{\phi}=(\phi_a)_a$, $\boldsymbol{\epsilon}(R)=\mathrm{diag}\{\epsilon_a(R)\}$, and
$\mathbf{F}_{ab}(\bv{R})=\langle \chi_{a}(R)|\boldsymbol{\nabla}_{\bv{R}}\chi_{b}(R)\rangle$
is the nonadiabatic coupling matrix. Equation~\eqref{ExcitonEqua1} contains both first-derivative couplings and the scalar term $\propto \mathbf{F}\!\cdot\!\mathbf{F}$ arising from the square. In general, no global unitary transformation eliminates $\mathbf{F}$, although local gauge choices can remove it in restricted regions. For numerical calculations we therefore adopt a diabatic (close-coupling) representation in which the kinetic-energy operator is diagonal and the couplings are collected into a potential matrix.

Defining a local unitary transformation $\mathbf{W}(\bv{R})$ that maps the adiabatic basis to a diabatic basis with vanishing derivative couplings, $\mathbf{F}'=0$, the stationary Schr\"odinger equation becomes
\begin{align}\label{ExcitonEqua2}
     \biggl[-\frac{\hbar^2}{2\mu}\boldsymbol{\nabla}^2_{\bv{R}}\,\mathbf{I}
     + \mathbf{V}(\bv{R})\biggr]\boldsymbol{\phi}'(\bv{R})
     = E\,\boldsymbol{\phi}'(\bv{R}),
\end{align}
where $\boldsymbol{\phi}'=\mathbf{W}^\dagger \boldsymbol{\phi}$ and $\mathbf{V}(\bv{R})=\mathbf{W}^\dagger(\bv{R})\,\boldsymbol{\epsilon}(R)\,\mathbf{W}(\bv{R})$ is the diabatic potential matrix~\cite{domcke2004conical}. Upon expanding $\phi_a'(\bv{R})=\sum_{lm}Y_m^{l}(\hat{\bv{R}})\,u_a^{l}(R)/R$, one obtains the coupled radial equations given as Eq.~(1) in the main text.

The coupled radial equations can be solved numerically on a discrete radial grid $R_0, R_1, \ldots, R_n$ by initializing the wavefunction at the short-range boundary and propagating outward to large $R$. The scattering matrix $S$ is then obtained by matching the solution at the last grid point to asymptotic $S$-matrix boundary conditions. In this work, we employ the renormalized Numerov method~\cite{karman2014renormalized} for stable propagation, which advances the solution via $\mathbf{Q}_i\,\mathbf{u}_i=\mathbf{u}_{i-1}$, where $\mathbf{u}_i$ and $\mathbf{u}_{i-1}$ denote the channel wavefunctions at consecutive grid points.
%
\subsection{S3.~ Short-Range Boundary Conditions and $\bv{S}$-Matrix Extraction} \label{sec;Bound}
%
To describe scattering in the multichannel setting, it is essential to impose appropriate short-range boundary conditions. In practice, when one is interested only in elastic and inelastic scattering, one often applies a hard-wall boundary condition at short range, $\bv{u}_0=0$, and propagates the coupled equations to large $R$. Here $\bv{u}_0$ denotes the wavefunction at the first grid point $R_0$. The hard-wall condition enforces an exponential decay at short range, consistent with the presence of a strongly repulsive potential barrier.

However, to capture reactive loss, where the Rydberg exciton is absorbed by the ionic impurity, a hard wall is insufficient because it neglects short-range reactive channels (which, microscopically, involve physics beyond the two-body description). In this work, we model reactivity by imposing a reactive boundary condition at short range and propagating the resulting complex-valued wavefunction to large $R$. This boundary condition accounts for the loss of probability flux into capture channels.

The flux-normalized universal reactive boundary condition is imposed in the adiabatic basis, where the interaction is locally diagonal and short-range capture is naturally defined channel by channel. This ties the universal loss condition directly to the local eigenchannels governing the short-range dynamics. For open channels it reads~\cite{janssen2012cold}
\begin{align}\label{eq:ShortBoundary}
    & \bv{u}_0 = \tilde{\bv{U}}_0\, \tilde{\bv{u}}_0, \CR
    &(\tilde{\bv{u}}_0)_{ab}
    = \sqrt{\frac{\mu k^{(0)}_{a}}{\hbar}}\, R_{0}\,
      h^{(1)}_{l_{a}}\!\left(k^{(0)}_{a} R_{0}\right)\delta_{ab},
\end{align}
where $\tilde{\bv{U}}_0$ is the orthogonal matrix that diagonalizes the potential matrix at $R_0$, and $\tilde{\bv{u}}_0$ is the wavefunction in the adiabatic basis. The Kronecker delta $\delta_{ab}$ ensures that the boundary condition is diagonal in channel space. The indices $a,b$ label the coupled channels and $l_a$ is the associated partial wave. The function $h^{(1)}_{l_a}$ is the spherical Hankel function of the first kind, representing purely incoming flux at $R_0$. The quantity $k^{(0)}_{a}$ is the local channel wavenumber at $R_0$, determined by the adiabatic potential energy $\epsilon^{(0)}_a$ and the total collision energy $E$. Channels with $\epsilon^{(0)}_a < E$ are open, while those with $\epsilon^{(0)}_a > E$ are closed.

For locally closed adiabatic channels, we use the exponentially decaying solution
\begin{align}
    (\tilde{\bv{u}}_0)_{ab} = e^{-k^{(0)}_a R_0}\,\delta_{ab}.
\end{align}

With the reactive boundary condition in Eq.~\eqref{eq:ShortBoundary}, we propagate the $\bv{Q}$ matrix outward to large $R$ and compute the $\bv{K}$ matrix as
\begin{align}
    \bv{K} = \bigl(\bv{G}_{n-1}-\bv{Q}_{n}\bv{G}_{n}\bigr)^{-1}\bigl(\bv{F}_{n-1}-\bv{Q}_{n}\bv{F}_{n}\bigr).
\end{align}
Here $\bv{F}$ and $\bv{G}$ are the regular and irregular reference solutions evaluated at the last two grid points. For open channels these are
\begin{align}
    (\tilde{\bv{F}}_{n})_{ab} &= \sqrt{\frac{\mu k^{(n)}_{a}}{\hbar}}\, R_{n}\, j_{l_{a}}\!\left(k^{(n)}_{a}R_{n}\right)\delta_{ab},\CR
    (\tilde{\bv{G}}_{n})_{ab} &= \sqrt{\frac{\mu k^{(n)}_{a}}{\hbar}}\, R_{n}\, n_{l_{a}}\!\left(k^{(n)}_{a}R_{n}\right)\delta_{ab},
\end{align}
where $j_l$ and $n_l$ are spherical Bessel functions of the first and second kind, respectively, and $k^{(n)}_a$ is the channel-resolved wavenumber at $R_n$. For closed channels, the regular and irregular solutions are given by the modified spherical Bessel functions:
\begin{align}
    (\tilde{\bv{F}}_{n})_{ab} &= k^{(n)}_{a}R_{n}\, i_{l_{a}}\!\left(k^{(n)}_{a}R_{n}\right)\delta_{ab},\CR
    (\tilde{\bv{G}}_{n})_{ab} &= k^{(n)}_{a}R_{n}\, k_{l_{a}}\!\left(k^{(n)}_{a}R_{n}\right)\delta_{ab}.
\end{align}

For open channels, the $\bv{S}$ matrix is obtained from the $\bv{K}$ matrix as
\begin{align}
    \bv{S} = (\bv{I}-i\bv{K})^{-1}(\bv{I}+i\bv{K}),
\end{align}
where $\bv{I}$ is the identity matrix. The channel-resolved cross sections for different scattering processes are obtained from the $S$~matrix as \cite{thompson2009nuclear}
\begin{align}\label{eq;CrossSection}
     \sigma^{(a)}_{\mathrm{el}} &= \frac{\pi}{k_a^2}\sum_{l}(2l+1)\,\big|1 - S^{aa}_{l}\big|^2, \CR
     \sigma^{(a)}_{\mathrm{in}} &= \frac{\pi}{k_a^2}\sum_{l}\sum_{b\neq a}(2l+1)\,\big|S^{ab}_{l}\big|^2, \CR
     \sigma^{(a)}_{\mathrm{re}} &= \frac{\pi}{k_a^2}\sum_{l}(2l+1)\,
        \left(1 - \sum_{b}\big|S^{ab}_{l}\big|^2 \right).
\end{align}
Here, $\sigma_{\mathrm{el}}^{(a)}$, $\sigma_{\mathrm{in}}^{(a)}$, and $\sigma_{\mathrm{re}}^{(a)}$ denote the elastic, inelastic, and reactive cross sections from the incoming channel $a$, respectively.

\begin{figure}[h!]
    \centering
    \includegraphics[width=0.5\columnwidth]{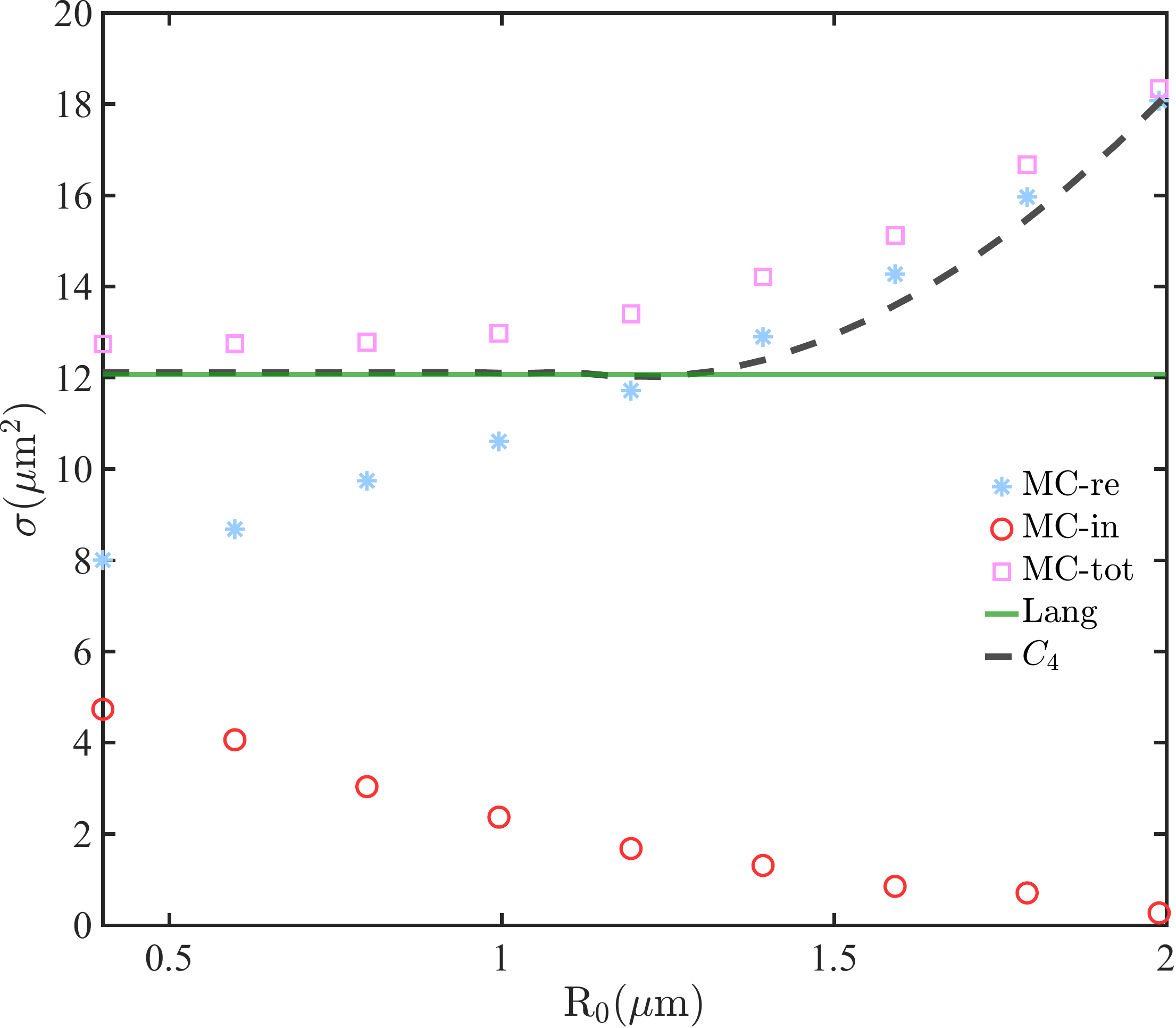}
    \caption{\textbf{Convergence of the cross sections:} Dependence on the absorbing-boundary radius $R_0$ at fixed energy $E = 24\,\mu\mathrm{eV}$. The reactive cross section (blue $\star$) decreases approximately linearly with reducing $R_0$, in contrast to the polarization-potential (single-channel) result (dashed black line), due to the emergence of an inelastic contribution (red $\circ$). The sum $\sigma_{\mathrm{re}}+\sigma_{\mathrm{in}}$ (magenta $\square$) converges for $R_0 \approx 0.75~\mu\mathrm{m}$, close to the LeRoy radius $R_L=0.73~\mu\mathrm{m}$.}
    \label{fig:boundary}
\end{figure}
To determine an appropriate placement of the absorbing boundary, we performed a convergence test by varying the absorbing radius $R_0$. We first consider the single-channel polarization potential, which describes the asymptotic interaction [Fig.~1(b) of the main article]. In this case, the reactive cross section converges to the Langevin capture cross section $\sigma_L$ (see \fref{fig:boundary}) when $R_0$ is chosen below a distance $R_{\max}=\frac{2\sqrt{C_4 M}}{\hbar\, l_{\mathrm{max}}}$ at which the effective potential attains its maximum value \cite{brouard2015tutorials}
\begin{align}
    V_{\mathrm{eff}}(R_{\max})=\frac{\hbar^4 l_{\mathrm{max}}^4}{16M^2 C_4},
\end{align}
where $l_{\mathrm{max}} = \left( \frac{16 M^2 C_4 E}{\hbar^4} \right)^{1/4}$ is the largest allowed partial waves that the exciton can overcome for a given collision energy ($E$).
For $R_0>R_{\max}$ and $l>l_{\max}$, the collision energy lies below the centrifugal barrier, so the motion is classically reflected at a turning point $R_t>R_{\max}$. Setting $R_t=R_0$, the cross section is given by 
\begin{align}
    \sigma_{C_4}(R_0)
    =
    \frac{1}{2}\sigma_L
    \left[
    \left(\frac{R_0}{R_{\max}}\right)^2
    +
    \left(\frac{R_{\max}}{R_0}\right)^2
    \right].
\end{align}

In contrast, for the full multichannel calculation, the reactive cross section does not exhibit the same convergence with $R_0$ due to redistribution of the flux into neighboring channels as shown in \fref{fig:boundary}. However, when the total loss, including both the flux removed by the reactive boundary condition and the flux transferred to adjacent channels, is considered, convergence is recovered close to the LeRoy radius $R_L$, defined as $R_L=2\sqrt{\langle r_n^2\rangle}$. Therefore, to focus on the physically relevant inelastic dynamics, we set the absorbing boundary at the corresponding LeRoy radius for the remaining principal quantum numbers.

%
\subsection{S4.~ Partial-Wave–Resolved Reactive–Inelastic Crossover} \label{sec;crossSec}
%
The cross section in the high-energy limit under single-photon excitation receives contributions from many partial waves. These partial waves exhibit different dynamics because the centrifugal barrier increases with $l$, which modifies how closely the exciton can approach the impurity and how strongly it samples the interaction region. Consequently, the kinetic energy varies across the scattering trajectory and depends on both $R$ and $l$. The radial kinetic energy in channel $a$ is
\begin{align}
    T^a_{r}(R;l)=E-\Delta E_{a}(R)-\frac{\hbar^{2}l(l+1)}{2MR^{2}},
\end{align}
where $\Delta E_{a}(R) = \epsilon_{a}(R)-\epsilon_{a}(R=\infty)$.  Asymptotically, $\Delta E_a(R) \to 0$ and the centrifugal term vanishes, so $T_r\to E$. As the exciton enters the interaction region, $T_r(R)$ is reshaped by the competition between the interaction potential (encoded in $\Delta E_{a}(R)$) and the centrifugal barrier, which suppresses access to short range.
\begin{figure}[h!]
    \centering
    \includegraphics[width=1.0\columnwidth]{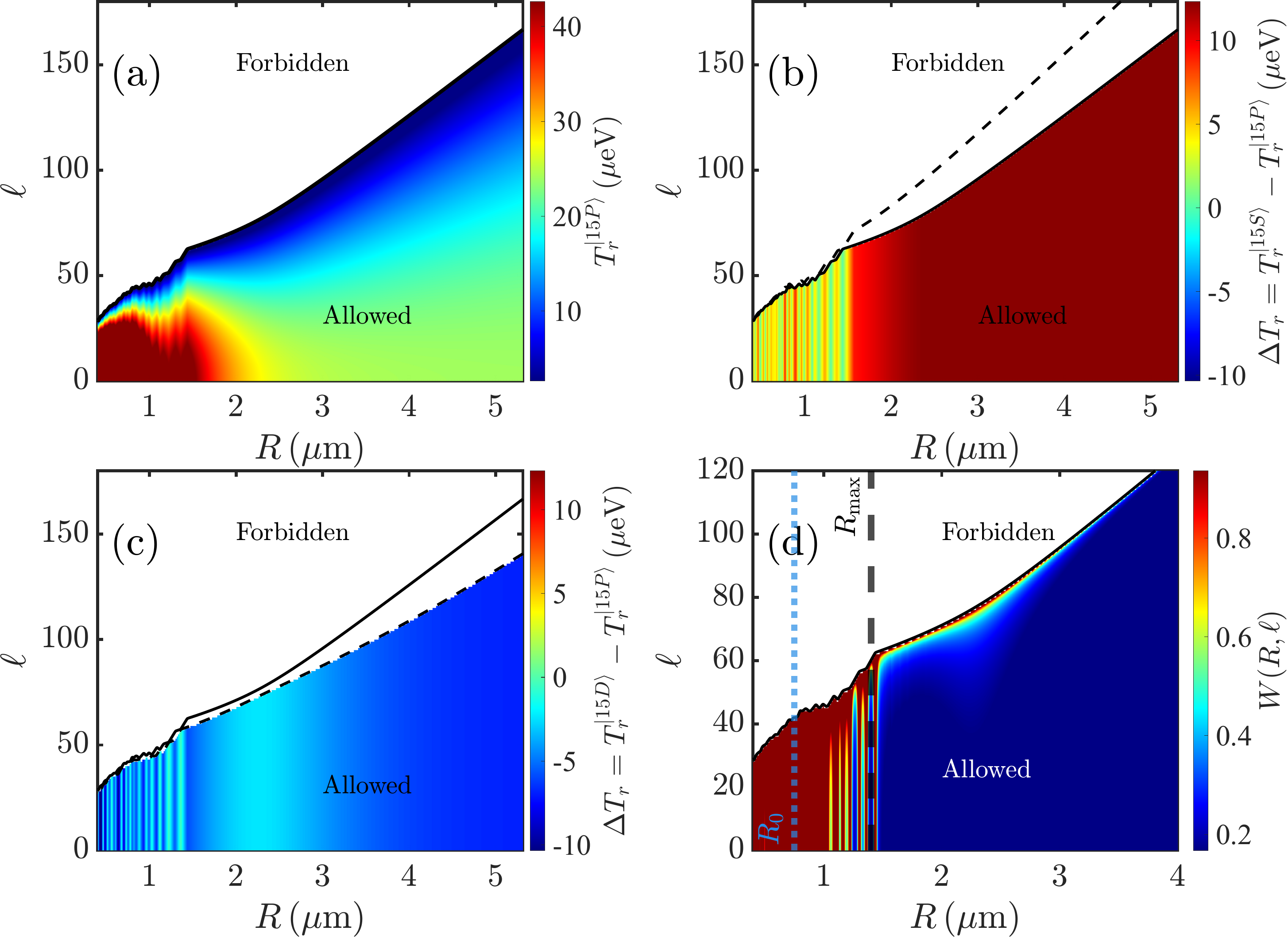}
    \caption{\textbf{Partial-wave--resolved exciton dynamics:} (a) Radial kinetic energy along the entrance-channel $\ket{15P}$ adiabatic surface for a fixed incoming kinetic energy $E = 24~\mu\mathrm{eV}$. (b) Change in kinetic energy relative to the entrance channel upon transfer to the lower-lying $\ket{15S}$ state, illustrating the short-range kinetic-energy gain. (c) Change in kinetic energy relative to the entrance channel upon transfer to the higher-lying $\ket{15D}$ manifold, illustrating the corresponding kinetic-energy reduction. (d) Velocity-scaled nonadiabatic coupling. The vertical dotted blue line indicates the absorbing radius, $R_0 = 0.73~\mu\mathrm{m}$, and $R_{\max} \approx 1.4~\mu\mathrm{m}$ (dashed black line) marks the distance at which the polarization potential reaches its maximum for $l_{\max}$. } 
    \label{fig:coupling}
\end{figure}

For small values of $l$, the exciton recoil energy overcomes the centrifugal barrier and allows access to smaller $R$, where $\Delta E_a(R)$ becomes increasingly attractive as $R$ decreases; as a result, $T_r$ increases, as shown in \frefp{fig:coupling}{a}. As $l$ increases, the centrifugal potential becomes comparable to the collision energy, causing a stronger reduction of $T_r$ at intermediate and short distances and thereby gradually suppressing the short-range kinetic energy. Beyond a critical $l$, $T_r$ vanishes at a classical turning point, so the motion is forbidden at smaller $R$ and the exciton is reflected before reaching the coupling region. Nonadiabatic coupling,
\begin{align}
    F_{ab}(R) = \frac{\bra{\chi_{a}}\partial_{R}H\ket{\chi_b}}{\epsilon_{b}(R)-\epsilon_a(R)},
\end{align}
enables transitions between adiabatic surfaces and thereby reshapes the kinetic-energy landscape. When the exciton transitions to a lower-lying manifold, the decrease in $\Delta E_a(R)$ is converted into additional kinetic energy, leading to a short-range acceleration, as shown in \frefp{fig:coupling}{b}. Conversely, transitions to higher-lying manifolds reduce $T_r$ and slow the motion (see \frefp{fig:coupling}{c}).

To quantify the $l$- and $R$-dependent transfer, we define a velocity-scaled nonadiabatic coupling,
\begin{align}
    W^{a}(R,l) = \frac{\sum_{b} F_{ab}(R)}{v^{a}_l(R)},
\end{align}
where $v^{a}_l(R)=\sqrt{2M T^a_r(R;l)}$ is the local velocity in channel $a$. As shown in \frefp{fig:coupling}{d}, $W^{a}(R,l)$ receives its dominant contribution from the region $R \lesssim R_{\max}$, indicating strong partial-wave-dependent interchannel coupling within the loss region defined by the polarization potential. For lower partial waves, trajectories that enter this region are also strongly accelerated and are therefore efficiently captured once they reach the reactive boundary at $R_0$. As the centrifugal barrier increases, the exciton slows down, and part of the incoming flux is transferred dominantly into higher-lying manifolds, where it subsequently reappears as inelastic scattering (See, Fig.~2 of the main article).

The inelastic contribution originates from nonadiabatic coupling in two distinct spatial regions. For $R \lesssim R_{\max}$, the reactive cross section is progressively converted into inelastic scattering up to the maximum partial wave $l_{\max}$ set by the long-range polarization potential. Consequently, the sum of the reactive and inelastic cross sections recovers the total loss predicted by the classical Langevin picture. By contrast, for $R>R_{\max}$, the nonadiabatic coupling produces additional inelastic transfer for partial waves $60 \lesssim l \lesssim 85$, as shown in \frefp{fig:coupling}{d}. This extra contribution causes the total loss, obtained by adding the reactive and inelastic cross sections, to deviate from the Langevin prediction. Thus, the extension of $W^{a}(R,l)$ to large partial waves promotes transfer into higher-lying channels at larger separations, diverting flux away from the short-range reactive region. Whether the flux is ultimately captured or reemerges in inelastic channels is therefore governed by the interplay between the spatial location of the coupling and the accessibility of the short-range reactive region.
\begin{figure}[h!]
    \centering
    \includegraphics[width=1.0\columnwidth]{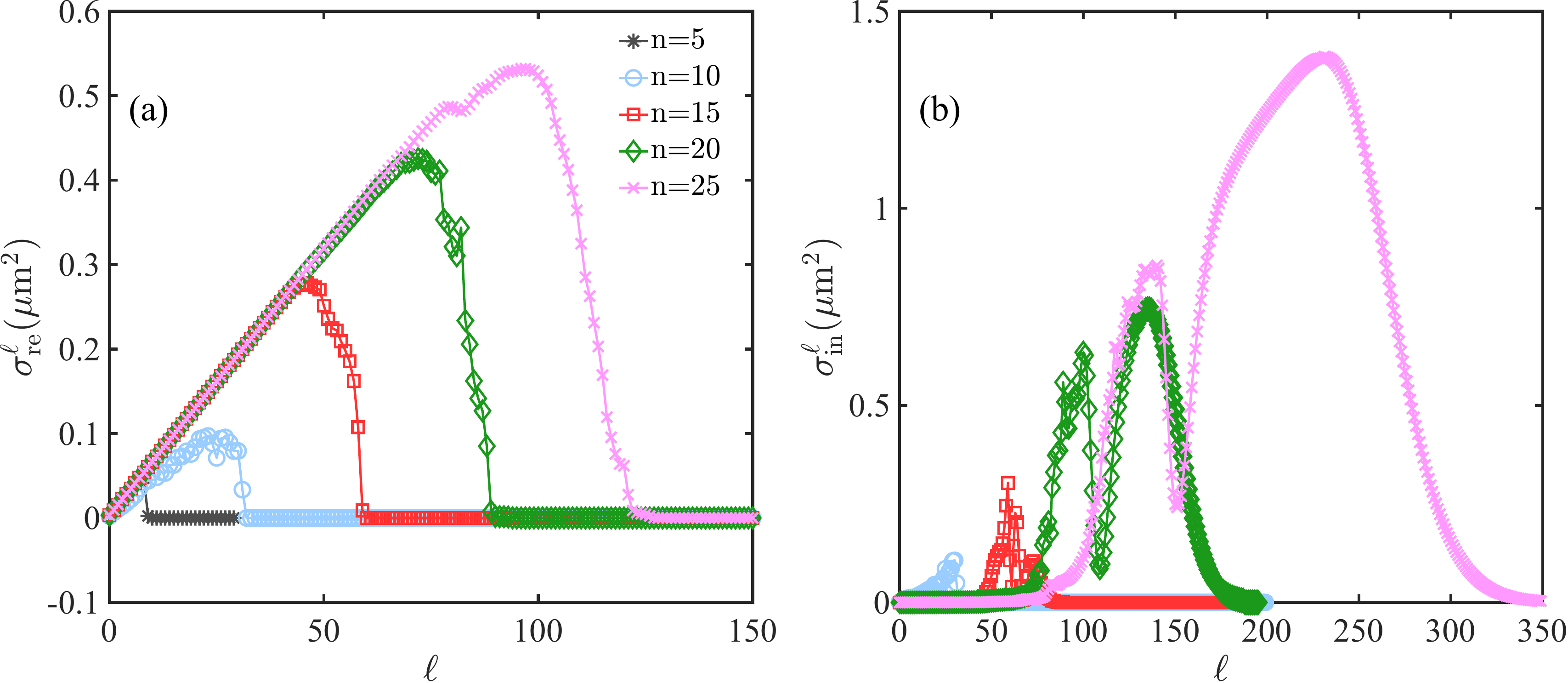}
    \caption{\textbf{Partial-wave cross sections at fixed collision energy $E = 24~\mu\mathrm{eV}$:} (a) The reactive cross section increases with principal quantum number $n$ as more partial waves contribute. (b) The inelastic contribution extends over a broader range of partial waves with increasing $n$, leading to a larger inelastic cross section. }
    \label{fig:cross}
\end{figure}

As the principal quantum number increases, the interaction (mixing) region shifts to larger separations. Consequently, a broader range of partial waves can access the nonadiabatic region without being excluded by the centrifugal barrier. This has two immediate consequences for the partial-wave-resolved cross sections. First, the reactive contribution extends to larger $l$, since more trajectories can still reach short range, as shown in \frefp{fig:cross}{a}. Second, once the centrifugal barrier suppresses direct short-range access, nonadiabatic coupling remains effective over a wide range of partial waves. As a result, reactive loss is progressively replaced by inelastic redistribution for $R \lesssim R_{\max}$, while outer-region coupling adds a further inelastic contribution from larger partial waves, as seen in \frefp{fig:cross}{b}. This explains the increasing deviation from the Langevin prediction at larger principal quantum numbers: the larger the deviation, the more important the outer-region coupling. In addition, the elastic cross section also increases with the principal quantum number, indicating reduced purification efficiency in the high-energy regime.

In contrast, for small $n$ the nonadiabatic coupling region is confined to short range. Only the lowest partial waves can access this region, while higher-$l$ trajectories are reflected by the centrifugal barrier before significant mixing occurs. As a result, the flux that reaches short range is predominantly captured, and inelastic transfer is strongly suppressed. Consequently, the partial-wave spectrum is dominated by reactive loss with negligible inelastic contribution (e.g., $n=5$ in \frefp{fig:cross}{b}).
%
\subsection{S5.~ Classical Scattering of Rydberg Exciton--Impurity} \label{sec;class}
%
 %
\begin{figure}[h!]
    \centering
    \includegraphics[width=1.0\columnwidth]{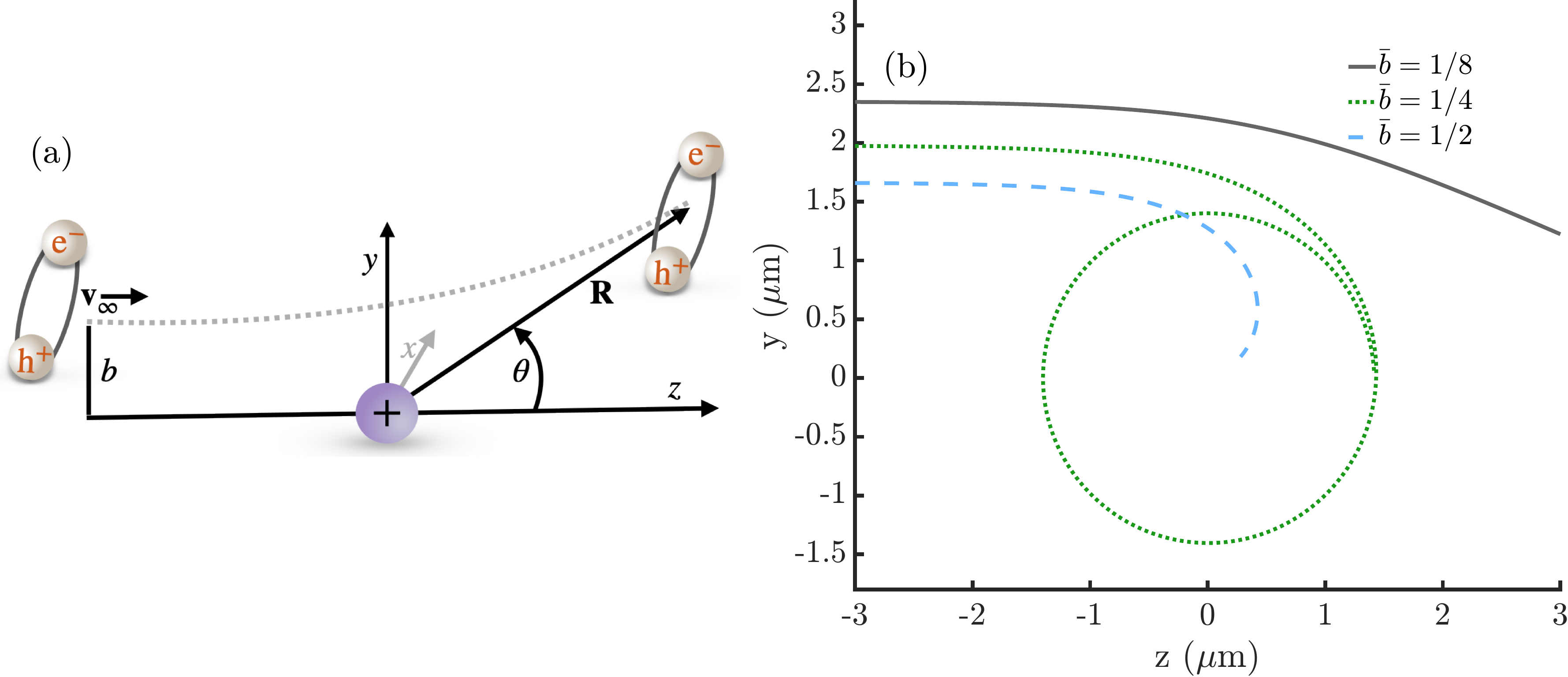}
    \caption{\textbf{Exciton--impurity scattering trajectories in the classical picture:} (a) An exciton approaches the impurity located at $\bv{R}=0$ along the positive $z$ axis with impact parameter $b$ (the perpendicular offset from the $z$ axis). In the asymptotic region the polarization potential vanishes, so the conserved total energy is purely kinetic. After the collision, the exciton again follows a straight-line trajectory at large $R$, defining the deflection (scattering) angle $\theta$ relative to the forward direction. (b) Representative trajectories obtained by solving Newton’s equation of motion in the polarization potential. At the critical value $\bar b=1/4$ (dotted green), the trajectory approaches an unstable circular ``orbiting'' motion of radius $R_{\mathrm{orb}}\approx 1.4~\mu\mathrm{m}$. For $\bar b=1/8$ (solid black), the exciton is reflected and escapes to infinity, whereas for $\bar b>1/4$ (e.g., $\bar b=1/2$, dashed blue) it penetrates to short range and is captured.
}
    \label{fig:class}
\end{figure}

The classical scattering of a Rydberg exciton from an charge impurity is governed by the long-range polarization potential $V(R)$. 
A schematic of the collision geometry is shown in \fref{fig:class}. 
The dynamics is described in terms of the relative coordinate $\bv{R}$, with the exciton approaching the impurity along the positive $z$-axis with asymptotic velocity $\bv{v}_{\infty}$. 
The time evolution of $\bv{R}(t)$ is determined by Newton’s equation of motion, $M\ddot{\bv{R}} = -\nabla V(\bv{R})$.

Since the interaction potential is radially symmetric, the angular momentum $\bv{L}=\bv{R}\times M\bv{v}$ is conserved. 
For the geometry shown in \fref{fig:class}, $\bv{L}$ points along the $x$-axis, and the scattering trajectory is therefore confined to the $y$--$z$ plane. 
The magnitude of the angular momentum is given by $L = Mbv_{\infty}$, where $b$ is the impact parameter. 
For a given collision energy $E = Mv_{\infty}^2/2$, the impact parameter is equivalently defined as $b = L/\sqrt{2ME}$.

With this choice of coordinates, the polar angle obeys $\dot{\theta} = -L/(MR^2)\le 0$, and decreases monotonically in time from its asymptotic value $\theta(t)\to\pi$ as $t\to -\infty$. 
As the particle approaches the scattering region from $R\to\infty$, energy conservation yields
\[
E = \frac{1}{2}M\dot{R}^2 + V_{\mathrm{eff}}(R),
\]
where the effective potential
\[
V_{\mathrm{eff}}(R) = V(R) + \frac{L^2}{2MR^2}
\]
governs the one-dimensional radial motion of the relative coordinate.

The scattering trajectory can be obtained by eliminating the time variable, giving
\begin{align}
    \frac{d\theta}{dR}
    = \frac{\dot{\theta}}{\dot{R}}
    = \pm \frac{L}{R^2 \sqrt{2M\bigl[E - V_{\mathrm{eff}}(R)\bigr]}},
\end{align}
where the $\pm$ signs correspond to the incoming and outgoing branches of the trajectory, respectively. 
Far from the scattering region ($R\to\infty$), the motion approaches a straight line with a small angular deviation from the forward direction. 
This deviation defines the deflection (scattering) angle $\Theta$, which is given by \cite{friedrich2015classical}
\begin{align}\label{eq:ClassPhase}
    \Theta(b)
    = \pi - \int_{R_\mathrm{ct}}^{\infty}
    \frac{2b}{R^2}
    \bigg[1-\frac{b^2}{R^2}-\frac{V(R)}{E}\bigg]^{-1/2}\,dR.
\end{align}

Here $R_{\mathrm{ct}}$ is the classical turning point, defined by the vanishing of the radial kinetic energy. 
We denote by $\Theta$ the scattering angle, to distinguish it from the polar coordinate $\theta$ used in the orbital equations. 
For the polarization potential, the turning point follows from $E=V_{\mathrm{eff}}(R)$, yielding
\begin{align}
R_{\mathrm{ct}}^{2}
&= \frac{L^{2}}{4ME}\left[1+\sqrt{1-\frac{16M^{2}C_{4}E}{L^{4}}}\right] \nonumber\\
&= \frac{b^{2}}{2}\left[1+\sqrt{1-\frac{4C_{4}}{Eb^{4}}}\right]
= \frac{b^{2}}{2}\left[1+\sqrt{1-4\bar b}\right],
\end{align}
where we introduced the dimensionless parameter $\bar b \equiv C_4/(E b^4)$. 
The turning point exists only for $\bar b\le 1/4$. 
At the critical value $\bar b=1/4$, the turning point merges with the top of the centrifugal barrier, corresponding to an (unstable) circular ``orbiting'' trajectory.
The radius of this orbit is $R_{\mathrm{orb}}=\left(\frac{C_4}{E}\right)^{1/4}$, as can be seen in \fref{fig:class}. 
For $\bar b>1/4$ (i.e., sufficiently small impact parameters), no classical turning point exists and the exciton penetrates to short range and is captured. 

The capture (Langevin) cross-section is obtained by counting all trajectories whose impact parameters allow them to surmount the centrifugal barrier.
From the condition $1-4\bar b=0$ one obtains the critical impact parameter $b_{\mathrm{abs}}=\left(\frac{4C_4}{E}\right)^{1/4}
\equiv \sqrt{2}\,R_{\mathrm{orb}}$. 
The Langevin cross-section is given as
\begin{equation}
\sigma_{\mathrm{L}}
=\pi b_{\mathrm{abs}}^{2}
=\pi\left(\frac{4C_4}{E}\right)^{1/2}
=2\pi\sqrt{\frac{C_4}{E}}.
\end{equation}

For $\bar b<1/4$, the centrifugal barrier dominates and the exciton is reflected at $R_{\mathrm{ct}}$. 
In the asymptotic limit $b\to\infty$, \eref{eq:ClassPhase} gives $\Theta(b)=\frac{3\pi C_4}{4Eb^{4}}$. 
More generally, for an attractive power-law potential $V(r)\sim -C_\alpha/r^\alpha$, the weak-scattering deflection angle scales as $\Theta(b)\propto C_\alpha/(E b^{\alpha})$ \cite{friedrich2015classical}. 

The differential cross-section follows directly from the deflection (scattering) angle, and is defined as the number of particles scattered into a solid angle $d\Omega=\sin\theta\,d\theta\,d\phi$ per unit incident flux,
\begin{align}\label{eq:diff}
    \frac{d\sigma}{d\Omega} &= \frac{b}{\sin{\theta}}\bigg|\frac{db}{d\theta}\bigg|
    = \frac{b}{\sin{\Theta}}\bigg|\frac{db}{d\Theta}\bigg|,\CR
    &= \frac{1}{4}\frac{1}{\sin{\Theta}} \sqrt{\frac{3\pi C_4}{4E}} \,\Theta^{-3/2},\CR
    &\overset{\Theta\rightarrow 0}{=} \frac{1}{4} \sqrt{\frac{3 \pi C_4}{4E}} \,\Theta^{-5/2},
\end{align}
where in the last step we used $\sin\Theta\simeq \Theta$ for small angles (and $\Theta$ coincides with the usual scattering polar angle $\theta$ for central-force scattering). 
Equation~\eqref{eq:diff} diverges in the forward direction due to the long-range nature of the polarization interaction.
Forward scattering ($\Theta\to 0$) is dominated by large impact parameters $b$, for which the trajectory experiences only a weak deflection.
Since the number of incident trajectories in an annulus $[b,b+db]$ scales as $2\pi b\,db$, large impact parameters dominate the phase space, leading to a forward-scattering divergence in the idealized long-range model. By contrast, the opposite limit $\Theta\to\pi$ probes a qualitatively different focusing mechanism: trajectories with a comparatively narrow range of impact parameters can concentrate into a backward cone, producing a ``glory-type'' enhancement \cite{adam2002mathematical}.

%
\section{S6.~ Backward Glory Interference in Elastic Exciton–-Impurity Scattering}
%
Backward glory scattering refers to an enhancement of the differential cross section near $\theta=\pi$, arising from axial focusing of classical trajectories. Unlike forward scattering, which is dominated by large impact parameters, the backward glory is governed by a finite range of impact parameters whose contributions interfere coherently in the backward direction, leading to a characteristic Bessel-type angular dependence \cite{berry1972semiclassical}.

For elastic scattering in channel $a$, the scattering amplitude and differential cross section are
\begin{align}
f_{aa}(\theta) &= \frac{1}{2ik}\sum_{\ell=0}^{\infty}(2\ell+1)\bigl(S_{aa,\ell}-1\bigr)P_\ell(\cos\theta),\\
\frac{d\sigma_{\mathrm{el}}}{d\Omega}(\theta) &= \bigl|f_{aa}(\theta)\bigr|^{2}.
\end{align}

To analyze the backward direction, we set $\Delta=\pi-\theta\ll 1$ and use
\begin{equation}
P_\ell(\cos(\pi-\Delta)) = (-1)^\ell P_\ell(\cos\Delta),
\end{equation}
which gives
\begin{equation}
f_{aa}(\pi-\Delta)=\frac{1}{2ik}\sum_{\ell}(2\ell+1)\bigl(S_{aa,\ell}-1\bigr)(-1)^\ell P_\ell(\cos\Delta).
\end{equation}

In the limit $\ell\gg 1$, $\Delta\ll 1$ with $(\ell+\tfrac12)\Delta$ fixed, the Legendre polynomial admits the approximation \cite{newton2013scattering}
\begin{equation}
P_\ell(\cos\Delta)\simeq J_0\!\bigl[(\ell+\tfrac12)\Delta\bigr],
\end{equation}
so that the backward amplitude becomes a coherent Bessel-weighted sum,
\begin{equation}
f_{aa}(\pi-\Delta)\simeq\frac{1}{2ik}\sum_{\ell}(2\ell+1)\bigl(S_{aa,\ell}-1\bigr)(-1)^\ell
J_0\!\bigl[(\ell+\tfrac12)\Delta\bigr].
\end{equation}

Introducing the semiclassical impact parameter $b=(\ell+\tfrac12)/k$, this expression can be interpreted as a coherent superposition of contributions from rings of radius $b$, each weighted by $J_0(kb\Delta)$. A backward glory arises when the dominant contribution comes from a narrow range of impact parameters near $b=b_g$, corresponding to axial focusing at $\theta=\pi$. In this limit,
\begin{equation}
f_{aa}(\pi-\Delta)\propto J_0(k b_g \Delta),
\qquad
\Rightarrow
\qquad
\frac{d\sigma_{\mathrm{el}}}{d\Omega}(\pi-\Delta)\propto J_0^{2}(k b_g \Delta).
\end{equation}
Consequently, the minima in the backward cone satisfy $k b_g \Delta_n=j_{0,n}$, where $j_{0,n}$ are the zeros of $J_0$. This implies the universal ratio relation
\begin{equation}
\frac{\Delta_n}{\Delta_1}=\frac{j_{0,n}}{j_{0,1}},
\end{equation}
which is independent of the overall normalization and therefore provides a robust diagnostic of backward-glory interference. The characteristic impact-parameter scale can be extracted directly from the first minimum,
\begin{equation}
b_g=\frac{j_{0,1}}{k\Delta_1}.
\end{equation}

\begin{figure}[h!]
    \centering
    \includegraphics[width=1.0\columnwidth]{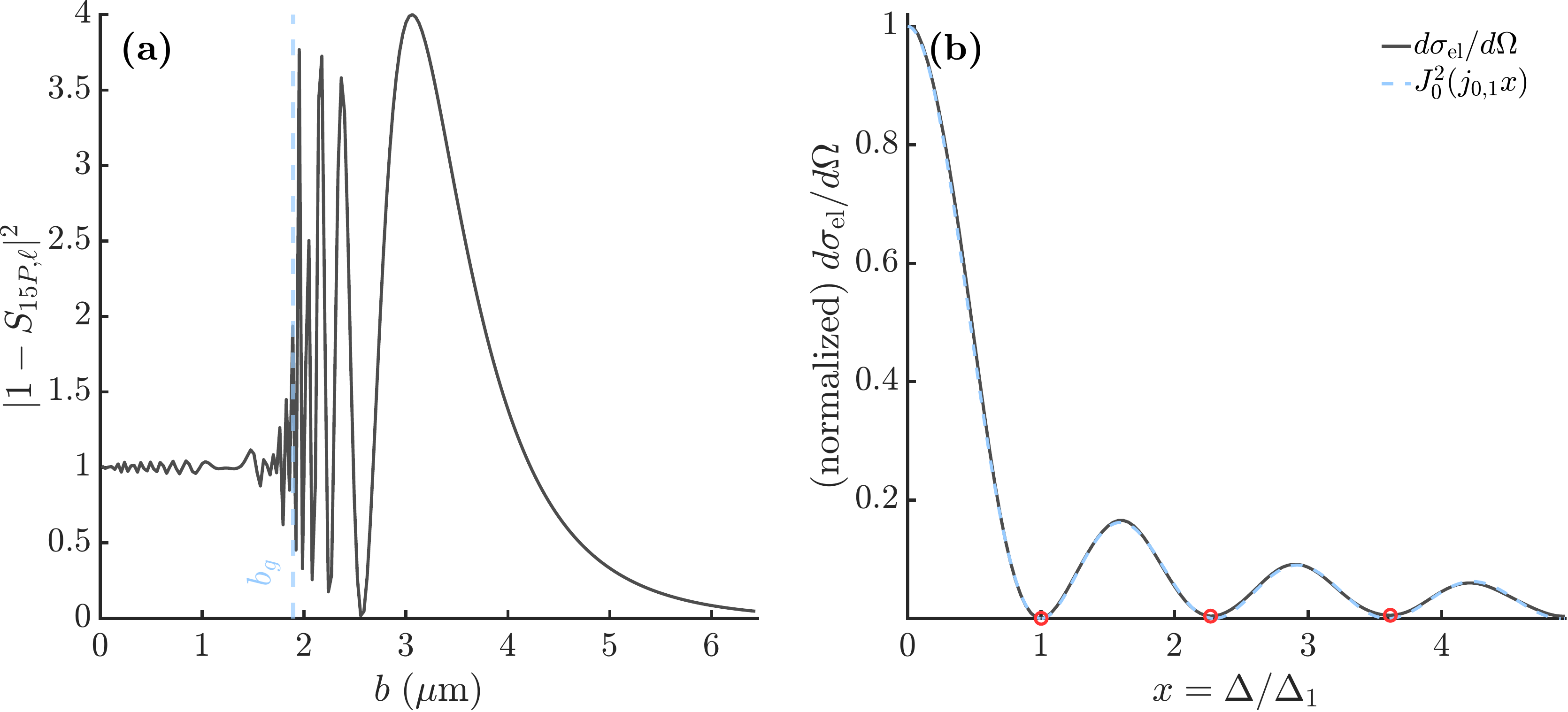}
    \caption{\textbf{Backward glory interference pattern:}
    (a) $|1-S_{15S,\ell}|^2$ (solid black) as a function of the semiclassical impact parameter $b$. The extracted glory scale $b_g\simeq 1.98~\mu\mathrm{m}$ lies close to the classical critical impact parameter $b_{\mathrm{abs}}$, indicating that a finite band of impact parameters near the capture threshold dominantly feeds the coherent backward amplitude.
    (b) Normalized elastic differential cross section in the backward cone (solid black) as a function of $x= \Delta/\Delta_1$, showing the semiclassical showing the semiclassical glory prediction (dashed blue). Here red $\circ$ indicates the minima, with $\Delta_1=0.0404$ is the first minimum and $k=31.5~\mu\mathrm{m}^{-1}$ is the wave number for the $\ket{15P}$ state.}
    \label{fig:glory}
\end{figure}

Figure~\ref{fig:glory}(b) demonstrates this backward-glory interference by collapsing the normalized elastic differential cross section onto the universal Bessel form: the data follow the predicted $J_0^2(j_{0,1}x)$ dependence and reproduce the corresponding sequence of minima. Figure~\ref{fig:glory}(a) relates the same angular pattern to the contributing partial waves by plotting $|1-S_{15P,\ell}|^2$ as a function of the semiclassical impact parameter $b=(\ell+\tfrac12)/k$, thereby highlighting the localized band of impact parameters that feeds the coherent backward amplitude. In multichannel scattering with absorption [nonunitary $S_{aa,\ell}$], loss primarily reduces the oscillation contrast through $|S_{aa,\ell}|<1$, but does not shift the positions of the Bessel minima provided that a coherent band of partial waves still contributes in the backward direction.

%
%

%

\end{document}